\documentclass[a4paper,fleqn,usenatbib]{mnras}
\usepackage[T1]{fontenc}
\usepackage{ae,aecompl}
\usepackage{color}
\usepackage{graphicx}
\usepackage{amsmath} 
\usepackage{amssymb}

\newcommand\blankline{\par\vskip 8pt\noindent}

\renewcommand\ion[2]{#1\,\,{\small\MakeUppercase{\romannumeral #2}}}

\newcommand{\nvres}{\ion{N}{5} $\lambda\lambda 1238,1242$}

\newcommand{\nivsev}{\ion{N}{4} $\lambda 1719$}

\newcommand{\nivuvtrip}{\ion{N}{4} $\lambda\lambda  3479 - 3485$}

\newcommand{\ciiiuvtrip}{\ion{C}{3} $\lambda\lambda  1175 - 1177$}
\newcommand{\oivdoub}{\ion{O}{4} $\lambda\lambda 3405 - 3414$}   %

\newcommand{\civopt}{\ion{C}{4} $\lambda \lambda 5801, 5812$}
\newcommand{\civres}{\ion{C}{4} $\lambda\lambda 1548,1552$}

\newcommand{\kms}{\hbox{km\,s$^{-1}$}}
\newcommand{\Vinf}{\hbox{$V_\infty$}}

\newcommand{\Rstar}{\hbox{$R_*$}}
\newcommand{\Rsun}{\hbox{$R_\odot$}}
\newcommand{\Msun}{\hbox{$M_\odot$}}
\newcommand{\Lsun}{\hbox{$L_\odot$}}
\newcommand{\Msunyr}{\hbox{$M_\odot \,\hbox{yr}^{-1}$}}
\newcommand{\Mdot}{\hbox{$\dot M$}}
\newcommand{\Teff}{\hbox{$T_{\rm eff}$}}

\newcommand{\cmfgen}{\textsc{cmfgen}}

\begin{document}

\title[BAT99-9 -- a WC4 Wolf-Rayet star with nitrogen emission]{BAT99-9 -- a WC4 Wolf-Rayet star with nitrogen emission:\\
Evidence for binary evolution?\thanks{This paper includes data gathered with the 6.5\,m Magellan telescopes located at Las Campanas Observatory, Chile.}}

\author[D. J. Hillier et al.]
{D. John Hillier,$^{1}$ 
Erin Aadland,$^{2,3}$
Philip Massey,$^{3,2}$ and  
Nidia Morrell$^4$ \\
$^{1}$ Department of Physics and Astronomy \& Pittsburgh Particle Physics, Astrophysics and Cosmology Center (PITT PACC), \\ \hspace{1cm}  University of Pittsburgh, 3941 O'Hara Street,  Pittsburgh, PA 15260, USA \thanks{hillier@pitt.edu} \\
$^{2}$ Department of Astronomy and Planetary Science, Northern Arizona
University,  Flagstaff, AZ 86011-6010, USA \\
$^{3}$ Lowell Observatory, 1400 W Mars Hill Road, Flagstaff, AZ 86001, USA  \\
$^{4}$ Las Campanas Observatory, The Carnegie Observatories,
Colina El Pino s/n, Casillas 601, La Serena, Chile}  

\date{Accepted XXX. Received YYY; in original form ZZZ}

\pubyear{2020}

\label{firstpage}
\pagerange{\pageref{firstpage}--\pageref{lastpage}}
\maketitle

\begin{abstract}
An analysis of the Large Magellanic Cloud (LMC) WC4 star BAT99-9 (HD 32125, FD 4, Brey 7, WS 3) shows that the star still contains photospheric nitrogen. Three N emission features (\nvres, \nivsev, \nivuvtrip) are unambiguously identified in the spectrum. CMFGEN models of the star yield a N/C ratio of  $0.004 \pm 0.002$ (by number) and a C/He ratio of $0.15_{-0.05}^{+0.10}$.  Due to the similarity of BAT99-9 to other WC4 stars, and the good fit achieved by CMFGEN to both the classic WC4 spectrum, and the N lines, we argue that the N lines are intrinsic to BAT99-9. An examination of a limited set of rotating models for single star evolution at LMC and Galactic metallicities shows that a model with a Galactic metallicity ($z=0.014$) and a progenitor mass of around 50\,\Msun\ can have a N/C ratio similar to, or larger than, what we observe for a significant fraction of its lifetime. However, the LMC models ($z=0.006$) are inconsistent with the observations. Both the single and  binary BPASS models predict that many WC stars can have a N/C ratio similar to, or larger than, what we observe for a significant fraction of their lifetime. 
While the binary models cover a wider range of luminosities and provide a somewhat better match to BAT99-9, it is not currently possible to rule out BAT99-9 being formed through single star evolution, given the uncertainties in mass-loss rates, and the treatment of convection and mixing processes.
\end{abstract}


\section{Introduction}

Classic Wolf-Rayet (WR) stars are the He-burning descendants of the most massive stars.   While it is generally accepted that WR stars are descended from O stars, their evolutionary channels are uncertain.  \cite{1967AcA....17..355P} suggested that they are formed as a result of binary stripping, which removes the outer hydrogen layers to reveal the nuclear processed material underneath.  On the other hand, \cite{1975MSRSL...9..193C} noted that this could also be achieved through mass loss via stellar winds.  The relative importance of the two mechanisms is a debate that continues to the present day  \cite[see discussion, for example, in][]{2019Galax...7...74N}.

The spectra of WR stars generally show strong and broad emission lines that  are formed in an optically thick stellar wind. Spectroscopically, they fall into three broad classes: WN, WC and WO. In the spectra of WN stars evidence for CNO processed material is seen. Many WN stars are devoid of hydrogen, and their surface layers typically
have a nitrogen abundance that exceeds the carbon and oxygen abundances by over a factor of 30
\cite[e.g.,][]{1995A&A...293..403C,2001ApJ...548..932H,2019A&A...627A.151S}. The WN class is very heterogeneous \citep[e.g.,][]{1983ApJ...268..228C}, most likely indicating that more than one channel contributes to
their formation. The spectrum of a WN star of a given spectral type (e.g., WN5) can exhibit narrow or broad emission lines and may, or may not,  show evidence for hydrogen. In this context, it is worth distinguishing H-rich WN stars from H-poor WN stars. Unlike most WR stars, which are core-He burning, many H-rich WR stars are still H-core burning, and appear as WR stars because they are very luminous and hence can generate a dense wind
\cite[e.g.,][]{2011A&A...535A..56G,2020MNRAS.491.4406S}. 

The  spectra of WO and WC stars are devoid of hydrogen lines, and show the products of He burning (e.g., $^{12}$C, $^{16}$O, $^{22}$Ne) at the stellar surface. As a group, the WC and WO stars are fairly homogeneous. There is a well-defined correlation, for example, between spectral type and line width \citep{1986ApJ...300..379T}. The WO and WC classes may not be distinct -- there is evidence that the properties vary smoothly between the two classes.
In WC stars the C mass fraction typically exceeds 0.1 \citep[e.g.,][]{HM99_WC,CDH02_WC,SHT12_GAL_WC}.

Since WC stars are more evolved than WN stars, it is expected that some WN stars will mature into WC stars. This evolution is known to be metal-dependent, and will also depend on the progenitor channel. An obvious question then arises -- are there stars intermediate between the WN and WC classes, and if so, are their numbers consistent with evolutionary predictions?

In a plot of the equivalent width (EW) of \civopt\ versus the EW of \ion{He}{2}\ $\lambda 4686,$ a handful of WN stars show anomalous (enhanced) \civopt\ emission \citep{1989ApJ...344..870M}.  These stars are generally referred to as WN/WC stars, with the suggestion that they are a transition between the WNs and WCs. Originally, these were thought to be WN+WC binaries (see, e.g., \citealt{1981SSRv...28..227V}). However, radial velocity studies of MR 111 and GP Cep showed that the C\,{\sc iv} and N lines moved in step \citep{1989ApJ...344..870M}.   \cite{1995A&A...304..269C} indicate the existence of nine transitional objects  within our galaxy, but it is unclear whether all these are true transitional objects or whether their composite nature is due to binarity. Another WN/WC star (WR 121-16) was recently found in our galaxy by \cite{2020ApJ...902...62Z} who deduce a nitrogen mass fraction of $1.5\pm1$\% and a carbon mass fraction of $0.2\pm0.1$\% using the PoWR atmosphere code \citep[e.g.,][]{2002A&A...387..244G,2015A&A...577A..13S}.

A classic example of this uncertainty is provided by HD 62910 (MR9, WR8) which is classified as WN6/WC4 \citep{1981SSRv...28..227V}  although \cite{1995A&A...304..269C} argue that the O and C line ratios are not consistent with the  WC4 classification. The spectrum  exhibits numerous carbon lines (e.g., \ion{C}{3} $\lambda$6740, \ion{C}{4} $\lambda 4658$, \ion{C}{4} $\lambda\lambda 5801, 5812$) as well as the classic nitrogen
lines (\ion{N}{4} $\lambda 4058$, \ion{N}{4} $\lambda\lambda 7103-7129$, \ion{N}{5} $\lambda\lambda 4604-4620$) \citep{1995A&A...304..269C}.  
\cite{1991IAUS..143..201N} presented evidence that the N and C lines moved in anti-phases, suggesting that the system was a binary.
A counter argument for binarity being the cause of the WN/WC spectrum was presented by \cite{1990A&A...232...89W} who found that both N and C lines showed the same correlation between line width and the excitation energy of the upper level.  Further, the line widths of the C lines are not consistent with those of other WC4 stars. A similar conclusion was reached by  \cite{1995A&A...304..269C} who found that its spectrum is consistent with that of a single star with abundances intermediate between WN and WC stars. In particular they found  N(C)/N(He)~$\simeq 0.02$, N(C)/N(N)~$ \simeq 3$, and N(C)/N(O)~$\simeq 4$. However, \cite{SHT12_GAL_WC} were unable to obtain a fit to the spectrum using a single star model.

Another example of a WN/WC star is WR 145 (MR 111). \cite{1990A&A...232...89W} suggest that it has a late O-type companion \citep[see also,][]{2009MNRAS.399.1977M}. After correcting for this star, they were generally able to obtain consistent fits to features in both the optical and IR. \cite{SHT12_GAL_WC} were also able to model both N and C
using a single star model, although they neglected the contribution by the companion O star
(Andreas Sander, private communication). To date, all stars bar one classified as WN/WC show  the strong nitrogen emission that exemplifies the WN spectral class. One possible exception
is WR126 which shows very strong carbon features in its optical spectrum, and relatively weak
\ion{N}{4} emission \citep{SHT12_GAL_WC}. However, as it is among the most luminous stars in the revised tabulation of WC and WN/WC luminosities by  \cite{2019A&A...621A..92S} it may be a binary system.

In this  paper we present evidence for the presence of nitrogen in the spectrum of the
Large Magellanic Cloud (LMC) WC4 star BAT99-9  (HD 32125, FD 4, Brey 7, WS 3) which indicates, in principle, that BAT99-9  belongs to the WN/WC subclass. However, BAT99-9 is very different from all other WN/WC stars that have been identified. BAT99-9 is clearly classifiable as a WC4 star, and its spectrum is virtually indistinguishable from other WC4 stars.  The detection of N in its spectrum is potentially an important diagnostic of how WC4 stars are created, the relevant time scales, and the possible importance of binarity.  

This paper is organized as follows. In \S\ref{Sect_obs}\ we discuss the observations utilized in this paper. The modeling  of BAT99-9 is presented in \S\ref{Sect_mod} while the evidence for the presence of nitrogen in BAT99-9 is presented in  \S\ref{Sect_nit}. We discuss the implications of our findings for the evolution of WR stars in \S\ref{Sect_conc}.

\section{Observations}
\label{Sect_obs}

Our BAT99-9 spectrophotometry extends from 1140\,\AA\ in the ultraviolet (UV)  through 2.5\micron\ in the near-IR (NIR). 

The UV observations were obtained with the Faint Object Spectrograph (FOS) on {\it HST} on UT 1994 September 26 using the 0\farcs26 diameter aperture and the G130H, G190H, and G270H gratings for a (total) wavelength coverage from 1140-3300\,\AA\ at a spectral resolving power $R=\lambda/\Delta \lambda$ of 1100-1600.  The exposure times were 1500~s, 650~s, 250~s, respectively.  The data sets were Y2A10203-6T, and were observed under program ID  5460 (PI: Hillier).   Standard pipeline reduced one-dimensional calibrated spectra were used in the analysis.

The optical data were taken on the Magellan 6.5-m Baade telescope with
the Magellan Echellette  (MagE; see \citealt{MagE}) and a 1\farcs0 slit. This gives a resolving power of $R=4100$, and covers the spectrum from 3150-9300\,\AA.  The spectrum was taken on UT 2016 January 11 under clear skies and 0\farcs67 seeing by N.I.M.  Three 900\,s exposures were made, with the star moved slightly along the slit between exposures in order
to reduce flat-fielding issues.   The slit was oriented to the parallactic angle, and the observation was taken at an airmass of 1.66.  Four spectrophotometric standards were observed, two at the start of the night and two at the end.
Wavelength calibration was by means of a ThAr arc.  The reduction process is described in detail in  \citet{2014ApJ...788...83M}.

The NIR spectra were also taken on the Baade telescope using the Folded port InfraRed Echellette (FIRE) using the 0\farcs75 slit that yielded a spectral resolving power of $R\sim5000$.  The spectrum was taken  on UT 2016 February 20 under clear skies and 0\farcs7-0\farcs9 seeing conditions, and consisted of four 600 s exposures in the standard A-B-B-A pattern used in the NIR.  Telluric correction and flux calibration were done using four 30 s exposures of HD~40750, an A0~V observed immediately after BAT99-9 and well-matched in airmass.  Wavelength calibration was by means of a ThAr lamp.
Data reduction was performed using the {\sc idl} FIRE pipeline kindly provided by the instrument PI, R. Simcoe and the 
FIRE team.

Both the optical and NIR BAT99-9 observations will suffer slit losses which may be greater than or less than those
of the standards used for flux calibration.  Therefore, we applied a single multiplicative factor (grey shift) to bring the optical data into accord with the FOS UV fluxes in the region of overlap, and then another single grey shift to bring
the NIR data into accord with the shifted optical data.  Of course, the regions of overlap are invariably in the regions where the instruments are at their poorest performance, and this correction process is uncertain at the 5\% level.  In general we adjust for this when fitting the spectra, but for this paper the adjustments are unimportant.

\section{Spectral Modeling}
\label{Sect_mod}

The modeling of BAT99-9 is part of a larger project to model Wolf-Rayet stars,  of types WC and WO, in the LMC. It is being undertaken using \cmfgen\ \citep{HM98_blank} which is a non-LTE radiative transfer code that was developed to model the atmospheres of stars with stellar winds. The free parameters in the model are the luminosity ($L$), the radius (defined at either $\tau_{\rm Ross=100}$ or at  $\tau_{\rm Ross=2/3}$), the abundances, the mass-loss rate (\Mdot), the velocity law [$V(r)$], and the volume-filling factor [$f(r)$]. Because of the high wind density, $\tau_{\rm Ross=2/3}$ is located in the wind ($V \sim 1000$\,\kms). Further the radius of the hydrostatic core is not well constrained by spectral modeling \cite[e.g.,][]{1991IAUS..143...59H,1997A&A...326.1117N,2004A&A...427..697H} --- its determination requires detailed hydrodynamic modeling. Calculations show that the iron opacity bump is crucial for driving the winds in WNE and WC stars \cite[e.g.,][]{NL02_mdot,2018A&A...614A..86G,2020MNRAS.491.4406S}.

For some models, we solved for the hydrostatic structure below the sonic point, in which case the effective temperature (\Teff), and surface gravity ($\log g$) are additional (or alternative) parameters. The hydrostatic structure is sensitive to the adopted opacities because of the Fe-group opacity peak, and because of the closeness of WC stars to the classical Eddington limit. However, because the wind is optically thick, the modeling is insensitive to the density structure below several hundred \kms. The luminosity, mass-loss rate, and abundances can be constrained without knowledge of this structure. Model parameters for BAT99-9, and a comparison with values derived in earlier studies, are presented in Table~\ref{Tab:params}. A detailed discussion of the models, and errors, will be given in a later paper.

For the comparisons between model and observations we have adopted a distance of 50\,kpc \citep{2019Natur.567..200P},  the reddening law of \cite{CCM89_ISE_fits}\  with $E(B-V)=0.08$ and $R=3.1$, and the
LMC reddening law of \cite{1983MNRAS.203..301H} with $E(B-V)=0.08$.

\begin{table*}
\caption{Stellar parameters for BAT99-9 and comparison with previously determined values.}
\begin{tabular}{cccccccccccl}
\hline
$\log L/\Lsun$ & \Teff$^1$  & \Rstar$^1$ & $\log \Mdot$ & $f$ & $\log \Mdot/\sqrt{f}$ & \Vinf & X(He) & X(C)$^2$ & X(N) & X(O) & Ref$^3$ \\
 & K & \Rsun& \Msunyr & &  \Msunyr & \kms \\
\hline
5.48  & 84000 & 2.6 & -4.85  &0.05&  $-$4.2& 2300& 0.66 & 0.29(0.14) & 0.0095(0.0004)&  0.041(0.015) & H20 \\
 5.44  &         &       &   &       &$-$4.3& 2500&          &     (0.13)  &  0                       &        (0.04)    & C02 \\
 5.29 &          &      & & & $-$4.2 &2300 &         &    (0.32)      &  0                     & (0.12)  & G98 \\
\hline
\\
\multicolumn{10}{l}{$^1$Defined at a Rosseland optical depth of 2/3.} \\
\multicolumn{12}{l}{$^2$Abundances are given in terms of the mass fractions. The values in parenthesis are the number ratios relative to helium.} \\
\multicolumn{10}{l}{$^3$References: H20  - current paper, C02 - \cite{CDH02_WC},  G98 - \cite{1998A&A...329..190G}.}  \end{tabular}

\label{Tab:params}
\end{table*}

\subsection{Evidence for nitrogen in BAT99-9}
\label{Sect_nit}

During the modeling of BAT99-9 it was noticed that a feature near 1720\,\AA\ was poorly fit. The feature shows a P~Cygni profile with absorption that extends to approximately 50\% below the continuum. 
We realized from the beginning that this feature bore a resemblance to the  \nivsev\  line found in WN stars; an examination of the NIST line list \citep{NIST571} reveals no other obvious candidate for this feature. It was subsequently realized that nitrogen could also be used to explain two other emission lines in BAT99-9: \nvres\ and \nivuvtrip.

\begin{figure}  
\includegraphics[width=1.0\linewidth, angle=0]{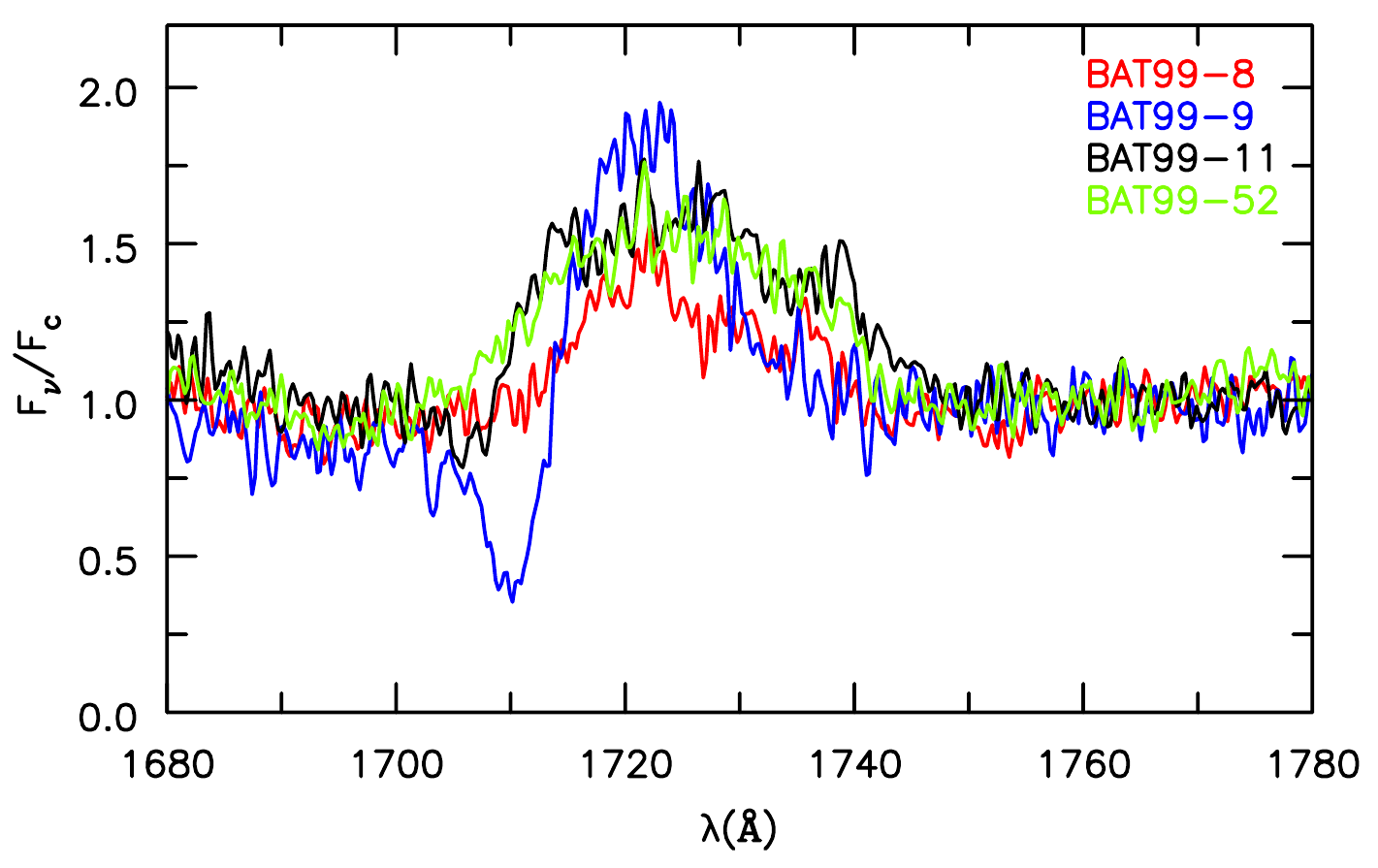}
\caption{Comparison of the spectral region containing the \nivsev\ transition. The spectra were normalized to unity using the region 1759 to 1770\,\AA. The presence of a strong blue-shifted absorption which we attribute to \nivsev\ is readily evident in BAT99-9, but is not evident in the other three stars. It is possible that BAT99-11 also contains a very weak blue-shifted absorption component, but the signal-to-noise is too low to make a definitive statement.   Other contributors to the emission around 1720\AA\ are Ne $\lambda 1718.06$ and \ion{S}{4}\ $\lambda\lambda 1723, 1727$.}
\label{fig_niv_comp}
\end{figure}

\begin{figure}  
\includegraphics[width=1.0\linewidth, angle=0]{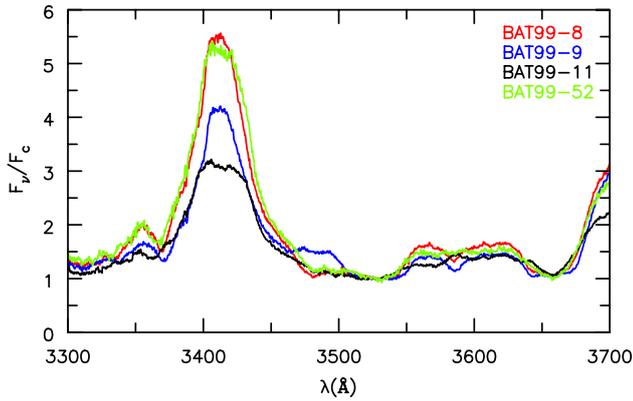}
\caption{Comparison of the region around the \nivuvtrip\ line. The spectra were normalized to unity using the region 3525 to 3530\,\AA. The presence of enhanced emission at 3480\,\AA\ in BAT99-9 is obvious, and can be attributed to the \nivuvtrip\ triplet. }
\label{fig_niv_trip_comp}
\end{figure}

A comparison of the  \nivsev\  profile in BAT99-9 with the same spectral region in three other WC4 stars is shown in Fig.~\ref{fig_niv_comp}.  Only in BAT99-9 does the emission feature near 1720\,\AA\ show a strong P~Cygni profile, although BAT99-11 may also exhibit a very weak absorption component. Additionally, only in BAT99-9 is there evidence for \nivuvtrip\ emission (Fig.~\ref{fig_niv_trip_comp}). 

\begin{figure}  
\includegraphics[width=1.0\linewidth, angle=0]{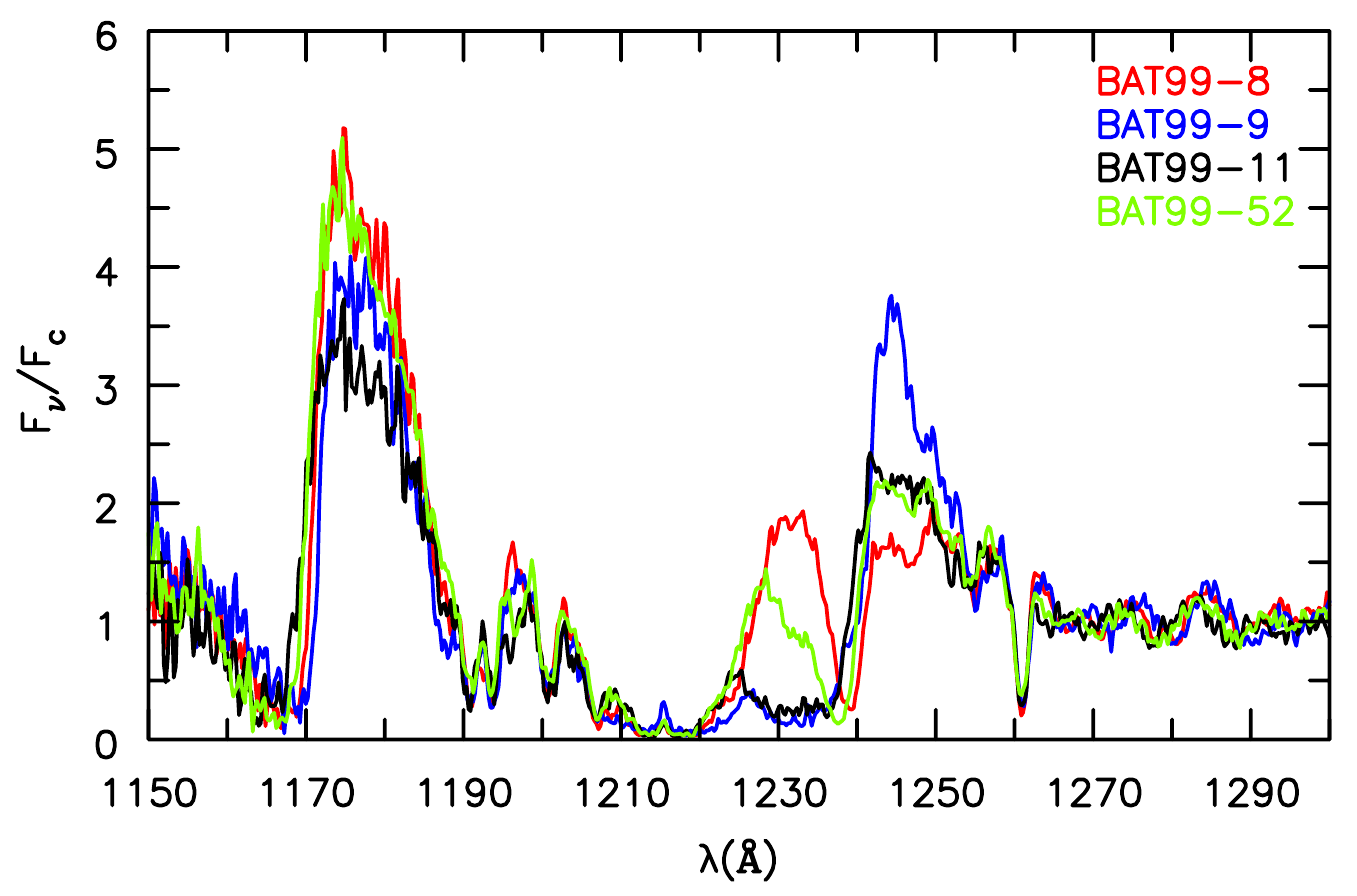}
\caption{Comparison of the region around the \nvres\ doublet. The spectra were normalized to unity using the region 1265 to 1270\,\AA. The presence of enhanced emission at 1240\,\AA\ in BAT99-9 is obvious, and can be attributed to the \nvres\ doublet. The extended blueshifted absorption that eats into the interstellar Ly$\alpha$ absorption profile is also readily evident. BAT99-11 may also have a \nvres\  contribution. Although it lacks the strong emission it does have extended blueshifted absorption that is very similar to BAT99-9. The strong emission centered near 1177\,\AA\  is \ciiiuvtrip. 
}
\label{fig_nv_comp}
\end{figure}

Determining whether the \nvres\ doublet is present in these four WC4 stars is more challenging. In BAT99-9 a very strong case for \nvres\ can be made -- there is enhanced emission at the correct wavelength and a strong and extended P~Cygni absorption indicating velocities approaching 3000\,\kms (which is similar to \civres)(Fig.~\ref{fig_nv_comp}). Two other stars, BAT99-8 and BAT99-52 show weak emission together with a narrower P~Cygni absorption component. However, in both cases the observed emission/absorption structure can be explained without invoking nitrogen. BAT99-11 is a little more interesting -- while its emission component is similar to BAT99-52 it shows an extended absorption component that is virtually identical to BAT99-9.  This may indicate that N is present in the atmosphere of BAT99-11, although the weakness of the emission component, and the weakness/absence of both \nivsev\ and \nivuvtrip, suggest that the abundance of N is less than in BAT99-9.

\begin{figure}  
\includegraphics[width=1.0\linewidth, angle=0]{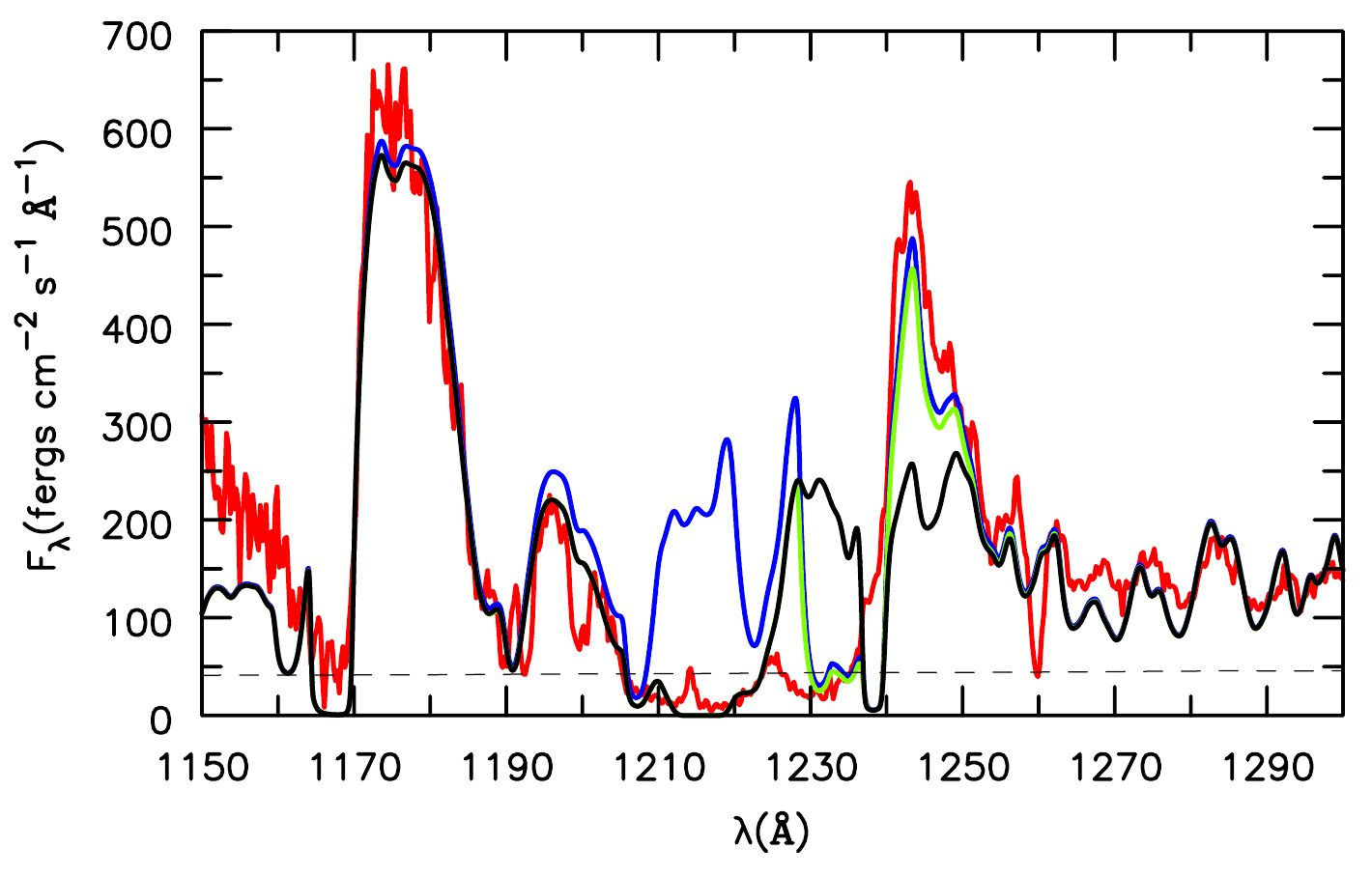}
\caption{Comparison of the \nvres\ doublet in BAT99-9 (noisy red curve) with the model (blue
and green curves).  The black curve shows the same spectrum but with N omitted from the spectral calculation. We have corrected the model spectrum for an interstellar H column density of 
$\log N_{\rm H}=21.3$ (green). The green curve is almost invisible in several places, since it overlaps closely with either the blue or black curve. The latter was also corrected for absorption by interstellar H.
 The broken line is the continuum associated with the model. The extent of the P~Cygni absorption in the model is somewhat smaller than in the observations. The same issue occurs with the \civres\ profile, and likely indicates that either \Vinf\ is somewhat too small, or that a larger microturbulent velocity (we
used 200\,\kms) in the outer wind is needed.  The same reddening corrections, and the same distance,
were used for this and subsequent plots.}
\label{fig_nv_mod}
\end{figure}

\begin{figure}  
\includegraphics[width=1.0\linewidth, angle=0]{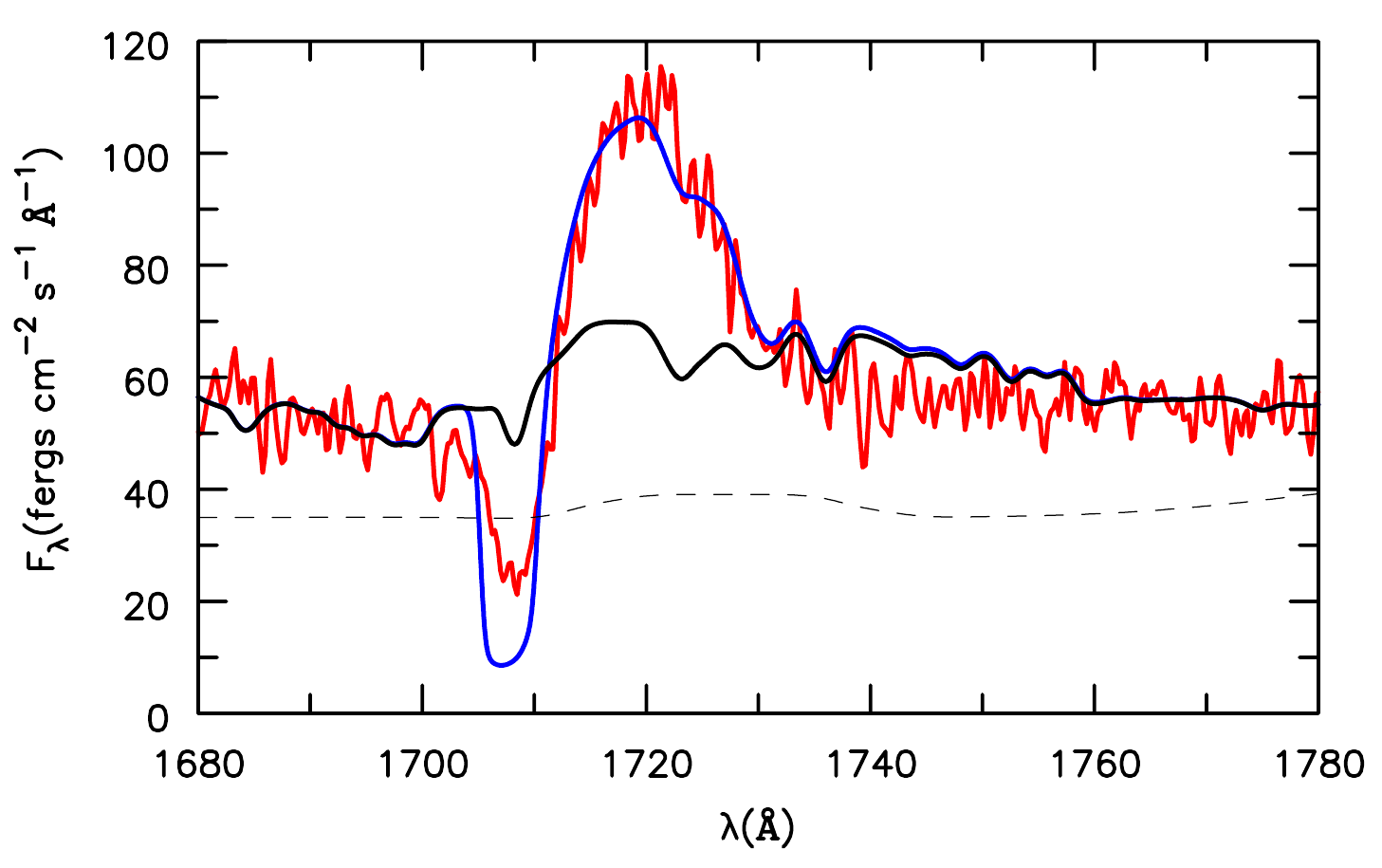}
\caption{Comparison of the \nivsev\ singlet in BAT99-9 (red) with the model (blue). The model
spectrum with nitrogen omitted from the flux calculation is also shown (solid black). The gross characteristics of the observed profile are reproduced by the model. The continuum (broken line) is not flat due to a resonance in one of the photoionization cross-sections.}
\label{fig_niv_mod}
\end{figure}

\begin{figure}  
\includegraphics[width=1.0\linewidth, angle=0]{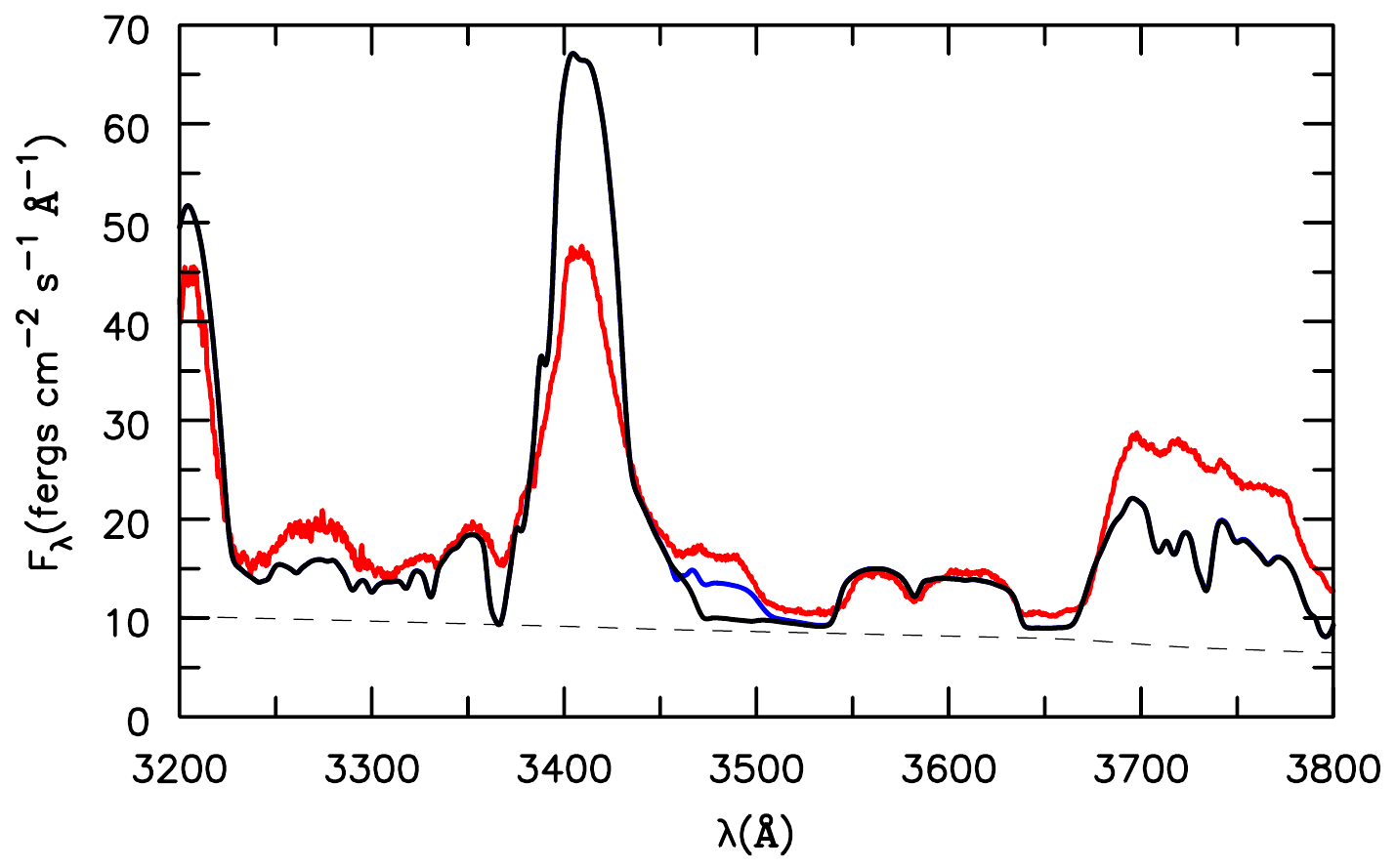}
\caption{Comparison of the \nivuvtrip\ multiplet with the model. The weak emission (just shortward of
3500\,\AA) associated with the \nivuvtrip\ multiplet (emphasized in blue) is not seen in the other WC4 stars (Fig.~\ref{fig_niv_trip_comp}). In our modeling, the strength of the \oivdoub\
multiplet (which is the strong emission feature just to the left of the much weaker \nivuvtrip\  emission)
is typically overestimated. }
\label{fig_nivopt_mod}
\end{figure}

For several reasons, it is unlikely that the nitrogen profiles in the spectra of BAT99-9 are due to the presence of a binary companion. First, and apart from the \ion{N}{5}  and \ion{N}{4} features alluded to earlier, the EWs of the emission  lines in BAT99-9 are similar to those in the other three WC4 stars -- BAT99-8, BAT99-11, and BAT99-52. Second, the \nivsev\ profile shows a classic P~Cygni profile, and since the P~Cygni absorption extends to $\sim 50$\% below the continuum, the WN star would need to contribute at least 50\% of the continuum light in the system (and likely more as the absorption component in the \nivsev\ profile in most WN stars does not reach ``zero''). This would then weaken the strength of the emission lines by at least a factor of two. This is inconsistent with the similarity of the emission line strengths in BAT99-9 to those seen in the other stars. Third, the absorption in the \nvres\ resonance doublet is close to zero whereas we might expect to see, in the presence of a companion star, a shelf on the profile. Such a shelf is seen, for example, in the double/quadruple system HD 5980 \citep[e.g.,][]{2011AJ....142..191G}.   Fourth, upon adding N to our WC4 models, it was found that the three nitrogen lines could be adequately matched using the best-fit WC model with a nitrogen mass fraction of 0.001, which is roughly a factor of 100 below the CNO equilibrium value reached during hydrogen burning (Figs. \ref{fig_nv_mod}, \ref{fig_niv_mod}, \ref{fig_nivopt_mod}). Finally, it is somewhat unlikely, although not impossible, that two stars in a binary system are both in the WR phase at the same epoch. The lifetime of a WC4 is strongly dependent on mass and the evolutionary channel leading to the WC stage. For BAT99-9, both the LMC single star models  \citep{2012A&A...537A.146E,GEM12_evol_wr}\footnote{The models we are using were kindly supplied by Georges Meynet and are for LMC metallicities (hereafter LMCSS). The model assumptions are similar to those used in the Galactic evolutionary models discussed by  \cite{2012A&A...537A.146E}.} and binary models \citep{2017PASA...34...58E} indicate that the remaining lifetime is less than $2 \times 10^5$ years.

\section{Discussion/Conclusion}
\label{Sect_conc}

Before examining the implications of our result for stellar evolution, we will discuss the variation of the N abundance with time as a consequence of stellar evolution. In massive stars, H burning proceeds primarily via the CNO cycle, and in the process much of the C and O is converted to N. Consequently,  in WN stars we obtain number ratios of  N(N)/N(C)$\,\,\sim 100$ and N(N)/N(O)$\,\,> 10$. In WN stars the nuclear material is revealed at the surface through mass loss and mixing processes.

During core He burning, $^{14}$N is rapidly converted to $^{22}$Ne via the capture of two alpha particles. As discussed by \cite{1991A&A...248..531L}, models that use the Schwarzschild criterion for convection predict a discontinuity between the He-rich layer containing N, and the CO layer.  As a consequence, we would not expect to see any stars that are transitioning between the WN and WC phases, or bona fide WC stars with nitrogen.  Several processes can change this simple picture.

In the models of  \cite{GEM12_evol_wr} non-rotating WR models that use the Schwarzschild criterion for convection do not, as expected, predict a WNC phase. As illustrated for a 60\,\Msun\ progenitor in their figure 6, the N abundance remains constant throughout most of the WN phase until it drops, precipitously, as the star enters the WC phase. On the other hand, for a rotating star there is a continuous and smooth transition from N(N)/N(C)$ > 1$ to  N(N)/N(C)$ < 1$ (their figure 6) -- 
a distinct WN/WC phase.  A change in the criterion for convection can also lead to a WN/WC phase.  When the Ledoux criterion is used, which allows for changes in radial composition, the efficiency of convection is altered, and this can lead to the appearance of some stars in the WN/WC phase [approximately 5\% if every WR transitions from WN to WC], and this is roughly consistent with observations \citep{1991A&A...248..531L}. 

Binary evolution can also lead to WN/WC transition stars \citep[e.g.,][]{2016MNRAS.459.1505M,2017PASA...34...58E}. Binarity influences the evolution of O stars and their descendants in multiple ways. For example, it affects mass-loss time scales since mass loss through the L1 Lagrangian point can be more efficient than via a stellar wind. It influences the lifetime of a WR star, and decreases the lower-mass limit of stars able to produce WR stars. In the BPASS LMC models the minimum mass of single stars that give rise to WC progenitors is 35\,\Msun (and for this mass the WC lifetime is only 1950\,yr), while the minimum mass
in binary systems is 12\,\Msun\ with a WC lifetime of  9540\,yr. Further, binarity influences rotation rates, and may directly influence mixing processes by tidal forces \citep[][and references therein]{2020A&A...641A..86H}.   

\begin{figure}  
\includegraphics[width=1.0\linewidth, angle=0]{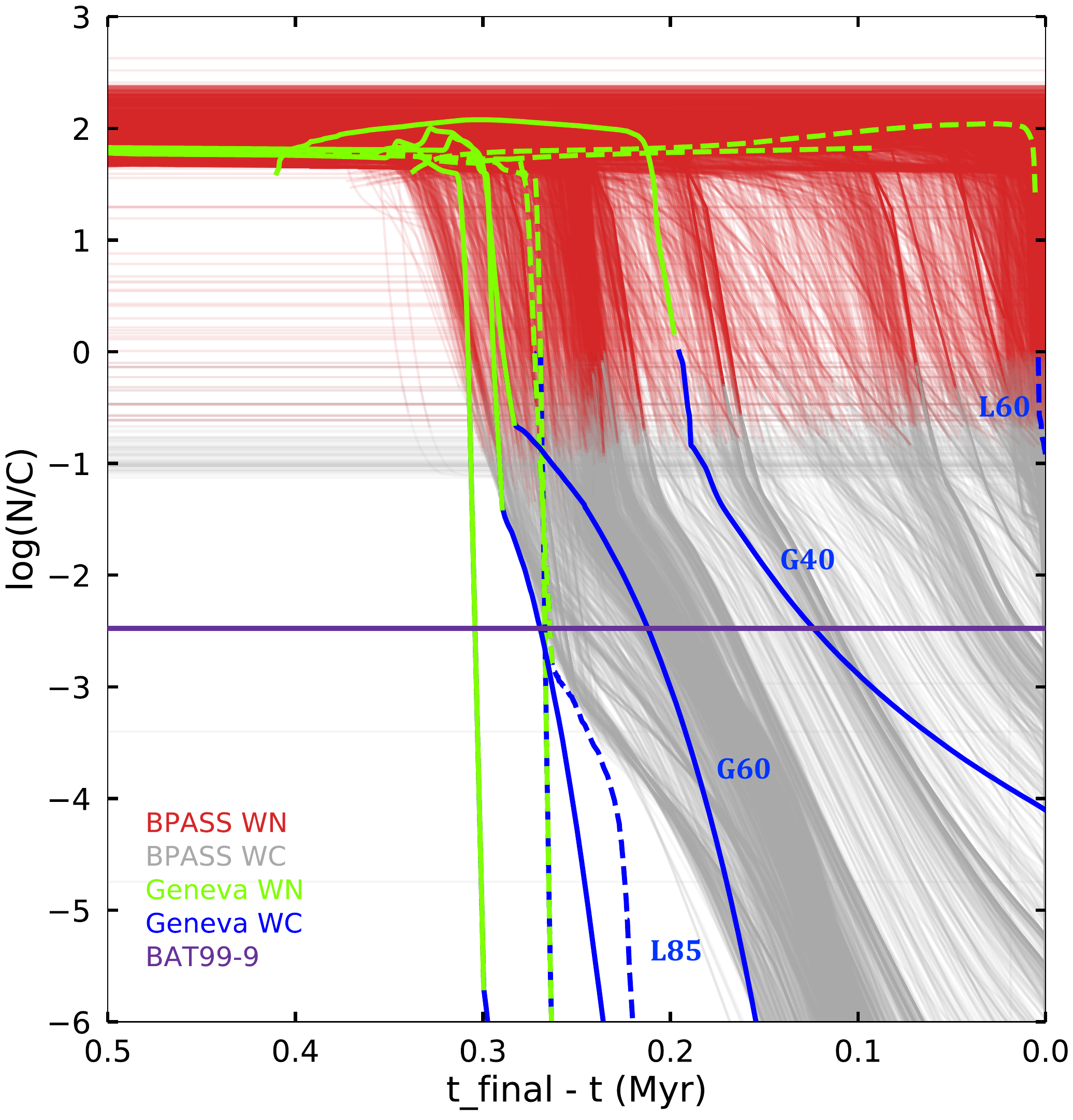}
\caption{Illustration of the variation of the N/C mass fraction ratio with time.  The LMC Geneva models are illustrated by dashed lines, while the solid lines are for Galactic models that assume an initial rotation velocity of 40\% of breakup. While many of the BPASS models can explain the observed N/C abundance ratio in BAT99-9, none of the LMC models are consistent. However, for the Galactic models a broad range of progenitor masses are ``consistent" with BAT99-9's N/C abundance ratio.  The progenitor masses of the Geneva models are 40\,\Msun, 60\,\Msun, 85\,\Msun, and 120\,\Msun. All of these models, except the 40\,\Msun\ LMCSS model, have a WC phase. However this model, and two \mbox{additional} LMCSS models that transition to the WC phase at a very low N/C abundance ratio (e.g., $10^{-10}$), are not shown. The purple line indicates the observed value for $\log$ N/C,  and several of the Geneva tracks are labeled by the progenitor mass.}\label{fig_t_NC}
\end{figure}

\begin{figure}  
\includegraphics[width=1.0\linewidth, angle=0]{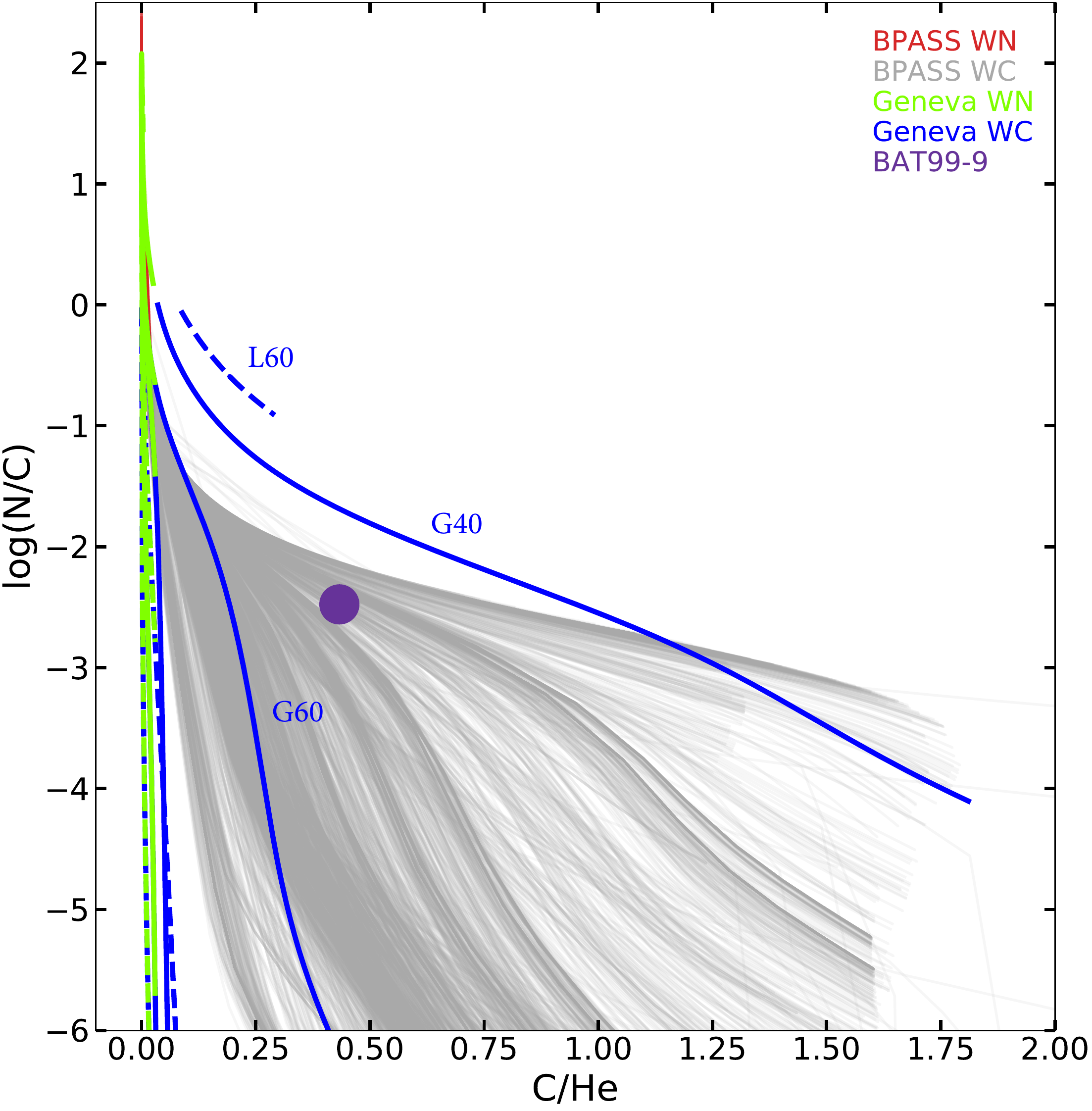}
\caption{Illustration of the variation of the N/C mass fraction ratio versus the C/He mass fraction ratios. A restricted set of the BPASS models can explain the observations of significant N in BAT99-9 at the observed C/He mass fraction ratio. A Galactic model with a progenitor mass near
\,50\Msun\ is consistent with the observations, but all LMC models are inconsistent.}
\label{fig_NC_Che}
\end{figure}

In Fig.~\ref{fig_t_NC}  we show the evolution of the X(N)/X(C) ratio as a function of time for the BPASS\footnote{Access to and  interpretation of the BPASS data was made with the assistance of {\sc hoki}, a software package used for making BPASS data accessible to the astronomical community
 \citep{2020JOSS....5.1987S}.} models \citep{2017PASA...34...58E,2018MNRAS.479...75S},   and the Geneva \mbox{LMCSS} and Galactic single star evolutionary models.  We utilize both sets of Geneva models due to uncertainties in the mass-loss rates that should be adopted in the evolutionary calculations, and which are one of the primary causes of the differences between the Galactic and LMCSS models. To further understand our results we show the X(N)/X(C) ratio as a function of the X(C)/X(He) ratio in Fig.~\ref{fig_NC_Che}.
 
The Geneva LMC single star models are inconsistent with the observations. A star with an initial mass of 40\,\Msun\ fails to become a WC star, and the duration of the WC phase for a 60\,\Msun\ progenitor is very short. Stars with higher mass can have a X(N)/X(C) ratio similar, or larger, than that observed, but for that time the models have a X(C)/X(He) ratio of less than 0.1 -- significantly smaller than the observed value of 0.4. Galactic single star models fare better --  a model with a progenitor mass of around 50\,\Msun\ gives results consistent with the observations.    

In general, the binary BPASS models can explain the observed X(N)/X(C) ratio. Typically the BPASS models have a X(N)/X(C) ratio greater than that observed for 10\% or more of the star's lifetime in the WC phase. Indeed, for some models, the X(N)/X(C) ratio is greater than that observed for a very significant fraction of their WC lifetime (e.g., 50\%). In addition, many BPASS models pass close to the location of BAT99-9 in the X(N)/X(C) versus X(C)/X(He) diagram. Some of the single star BPASS models are also consistent with the observed abundances ratios.  However, these models are more luminous than BAT99-9 -- by 0.25 dex or more. The differences between the Geneva and BPASS single star models highlight the considerable uncertainty still associated with theoretical stellar evolution models for massive stars.

The current evolutionary models indicate that binary evolution models are somewhat better able to explain the observed X(N)/X(C) and X(C)/X(He) ratios in the LMC WC star BAT99-9 and its luminosity.  However a limitation of our study is that we only compare our results to a limited set of evolutionary models that have been computed using a restricted set of parameters and assumptions. In particular, the models do not use the Ledoux criterion for convection and this will affect the reduction in the N abundance as the star enters the WC phase. Further, for example, the Geneva models are only calculated for two zero-age main sequence (ZAMS) rotation rates (0 and 40\% of critical) whereas some O stars will potentially rotate at even faster rates on the ZAMS. Finally, as noted earlier,  the mass-loss rates used in evolutionary models have significant uncertainties -- different mass-loss rates, for example, will alter the evolution of the star in the Hertzsprung-Russell diagram and its observed abundances.
 
As noted earlier, the spectrum of BAT99-9 is very similar to the other WC stars in our sample, and there is no evidence that a companion star influences the observed spectrum. A close examination of the spectrum of  BAT99-9 reveals that H$\gamma$ at 4340\,\AA\ is not present.  Thus  any companion star must be at least a factor of five fainter in the continuum at B than the WC star. With higher signal-to-noise spectra it will be possible to improve this limit -- in BAT99-9 the blended emission feature at 4340\,\AA\ appears to have a smooth rounded profile.

BAT99-9 is the first confirmed WC star with intrinsic N emission, and as such it cannot be used to make strong statistical inferences about the fraction of WC stars still possessing detectable N in their atmospheres. However, a future re-examination of archival optical and UV spectra will allow such inferences, and is encouraged. The three N lines seen in  the present study (\nvres; \nivsev; \nivuvtrip) do offer a ready means to probe for N in WC stars. In the hotter WC stars, the \nvres\ resonance transition will offer the most sensitivity for the detection of N. Indeed, our own observations suggest that the atmosphere of  BAT99-11 may also contain N, albeit with a lower mass fraction than BAT99-9. It is also apparent from Figs. \ref{fig_t_NC}  and  \ref{fig_NC_Che} that many WC stars may possess a measurable N abundance. Measurements of the N/C ratio in WC stars have the potential to place important constraints on the progenitors of WC stars, and/or provide important constraints on physical processes such as convection and mixing.

\blankline
{\bf Acknowledgements}
The authors thank the referee for their very careful reading of the manuscript, and their thoughtful comments. We also thank Dr. Heloise Stevance for assisting one of us (Erin Aadland) with the BPASS models and for providing thoughtful comments on a draft of this paper, Georges Meynet for providing useful insights during this work, and Cyril Georgy who supplied us with the LMC Geneva models. This work made use of v2.2.1 of the Binary Population and Spectral Synthesis (BPASS) models as described in \cite{2017PASA...34...58E}  and \cite{2018MNRAS.479...75S}. UV data presented in this paper were obtained from the Mikulski Archive for Space Telescopes (MAST).  Support for MAST for non-HST data is provided by the NASA Office of Space Science via grant NNX09AF08G and by other grants and contracts. Partial support to DJH for this work was provided by  STScI grants  HST-GO-13781.002-A and HST-AR-14568.001-A, and through NASA grant 80NSSC18K0729. STScI is operated by the Association of Universities for Research in Astronomy, Inc., under NASA contract NAS 5-26555.  EA and PM were provided funding through STScI grant HST-GO-13781.001, and through NASA grant 80NSSC18K0729; their ground-based observations were supported through the National Science Foundation AST-1612874.   Observing time at Las Campanas Observatory was obtained through the Carnegie Observatory's and Steward Observatory's Arizona Telescope Committees. The authors are grateful for their support, and for the excellent support received from the staff on the mountain.  This research has made use of NASA's Astrophysics Data System.

\blankline{\bf Data Availability}
The UV data used in this study is available via the HST data archive. Optical data will be made available on request. An earlier version of \cmfgen, and the associated atomic data, is available at
www.pitt.edu/\verb+~+hillier. Updates of this website are routinely made.  Models underlying this article will be shared on reasonable request to the corresponding author.

\bibliographystyle{mnras}
\bibliography{IAU250} 

\bsp
\label{lastpage}
\end{document}